\documentclass[10pt]{article}
\usepackage[top=1in,left=1.25in]{geometry}
\usepackage[affil-it]{authblk} % author affiliations
\usepackage{cite, amsmath, amssymb}
\usepackage{url} 
\usepackage{graphicx}
\usepackage{enumitem}
\usepackage{multirow}
\usepackage{xcolor,colortbl}

\begin{document}
	
\title{Dynamical compensation and structural identifiability: analysis, implications, and reconciliation}
\author[1,*]{Alejandro F. Villaverde}
\author[1]{Julio R. Banga}
\affil[1]{Bioprocess Engineering Group, IIM-CSIC, Vigo, Spain}
\affil[*]{afvillaverde@iim.csic.es}

\maketitle
	
%%%%%%%%%%%%%%%%%%%%%%%%%%%%%%%%%%%%%%%%%%%%%%%%%%%%%%%%%%%%%%%%%%%%%%%%%%%%

\begin{abstract}
The concept of dynamical compensation has been recently introduced to describe the ability of a biological system to keep its output dynamics unchanged in the face of varying parameters. 
Here we show that, according to its original definition, dynamical compensation is equivalent to lack of structural identifiability. This is relevant if model parameters need to be estimated, which is often the case in biological modelling.  
This realization prompts us to warn that care should we taken when using an unidentifiable model to extract biological insight: the estimated values of structurally unidentifiable parameters are meaningless, and model predictions about unmeasured state variables can be wrong. 
Taking this into account, we explore alternative definitions of dynamical compensation that do not necessarily imply structural unidentifiability. Accordingly, we show different ways in which a model can be made identifiable while exhibiting dynamical compensation. Our analyses enable the use of the new concept of dynamical compensation in the context of parameter identification, and reconcile it with the desirable property of structural identifiability.

\end{abstract}

%%%%%%%%%%%%%%%%%%%%%%%%%%%%%%%%%%%%%%%%%%%%%%%%%%%%%%%%%%%%%%%%%%%%%%%%%%%%	

\flushbottom
\maketitle
\thispagestyle{empty}

%%%%%%%%%%%%%%%%%%%%%%%%%%%%%%%%%%%%%%%%%%%%%%%%%%%%%%%%%%%%%%%%%%%%%%%%%%%%%%%%%%%%%%%%%%%%%%%	
\section*{Introduction}\label{sec:intro}

Some biological systems are capable of maintaining an approximatively constant output despite environmental fluctuations. The ability to keep a constant \textit{steady state} has been called homeostasis or exact adaptation, a feature that is known to be achievable with integral feedback \cite{Barkal,alon1999robustness,stelling2004robustness,ma2009defining,briat2016antithetic}. 
The ability of preserving not only the steady state, but also the \textit{transient response} (i.e. the dynamic behaviour) has been less studied and, despite recent contributions \cite{karin,young2017dynamics}, the mechanisms that make it possible are still less well understood.
%although recent work is correcting this \cite{karin,young2017dynamics}.
To account for this property, Karin et al. \cite{karin} have recently introduced the concept of dynamical compensation, which they defined as follows: 
%\textbf{DC1 definition of dynamical compensation:}
``Consider a system with an input $u(t)$ and an output $y(t,s)$ such that $s > 0$ is a parameter of the system. The system is initially at steady state with $u(0) = 0$. Dynamical compensation (DC) with respect to $s$ is that for any input $u(t)$ and any (constant) $s$ the output of the system $y(t,s)$ does not depend on $s$. That is, for any $s_1$, $s_2$ and for any time-dependent input $u(t)$, $y(t,s_1) = y(t,s_2)$''. Karin et al. \cite{karin} used this this definition of dynamical compensation, which we refer to as ``DC1'', to describe a design principle that provides robustness to physiological circuits.

%%% DC and SUI: 
The DC1 property is similar to the classic definition of structural unidentifiability. A parameter is structurally unidentifiable if it cannot be determined from any experiment, because there are different parameter values that produce the same observations. Using the same notation as in the DC1 definition, structural unidentifiability can be defined as follows \cite{ljung1994global,walter1997identification}: a parameter $s$ is structurally identifiable if it can be uniquely determined from the system output, that is, if for any $s_1$, $s_2$ it holds that $y(t,s_1) = y(t,s_2) \Leftrightarrow s_1 = s_2$. If this relationship does not hold for any $u(t)$, even in a small neighbourhood of $s$, the parameter is structurally unidentifiable. Thus, DC1 can be considered as a particular case of structural unidentifiability, with the additional requirement that the system is initially at steady state and with zero input \cite{sontag2016dynamic,villaverde2017dynamical}.

%%% SI:
Structural identifiability (SI) is a well-established concept with a long history of applications in the biological sciences \cite{bellman1970structural}. It was developed primarily by researchers working on the interface between biology and systems and control theory%  \cite{bellman1970structural,pohjanpalo1978system,cobelli1980parameter,vajda1989similarity,vajda1989state}
, long before the term ``systems biology'' became popular \cite{distefano2014dynamic}. %The topic has seen many advances in the last decades  \cite{evans2000extensions,evans2002identifiability,yates2009structural,bellu2007daisy,sedoglavic2002probabilistic,anguelova2012efficient,xia2003identifiability,meshkat2009algorithm,raue2009structural,stigter2015fast,villaverde2016}, and a number of critical evaluations of the available methods have been published recently \cite{Chis11a,miao2011identifiability,grandjean2014structural,raue2014comparison,villaverde2016identifiability}.
Beginning with its original conception, it was recognized that structural identifiability is concerned with the theoretical existence of unique solutions and therefore is, strictly speaking, a mathematical \textit{a priori} problem  \cite{jacquez1985numerical}. Solutions to \textit{a priori} problems depend only on the model’s structure (that is, the set of differential equations describing the dynamics of the system -- with the dependence on the inputs -- and the observation function that defines the measured variables) and not on the quality or quantity of the measurements the model is endeavouring to describe. This means that \textit{a priori} problems can be interrogated and completely understood before performing any experiments. Such interrogation is important: choosing a model structure is always based on arbitrary decisions that can have functional consequences, so testing structural properties allows us to detect defects in the chosen model structure before conducting further analysis \cite{walter1997identification}.
For further information about the available methodologies for structural identifiability analysis we refer the reader to recent reviews and comparisons \cite{Chis11a,miao2011identifiability,grandjean2014structural,raue2014comparison,villaverde2016identifiability} and references therein.

%%% Remarks on the Parameter Identification Problem in Systems Biology (In the Introduction, or in a separate section?
It should be noted that DC1 and structural unidentifiability are equivalent if we adopt a terminology commonly used in biological modelling, according to which the word ``parameter'' means a constant whose value is in general unknown \cite{distefano2014dynamic}. If all the parameters are assumed to be known, this equivalence does not apply because parameters are not being identified. However, in practice models usually contain a number of unknown parameters, whose values need to be determined. Even in those cases when parameter values can be obtained from the literature, they usually need to be reconciled with experimental data by some identification procedure based on input-output data \cite{van2006dynamic,jaqaman2006linking,ashyraliyev2008systems}. Thus, in realistic situations the correspondence between dynamical compensation and structural unidentifiability is relevant.
%%% The issue with SUI:
In this context, lack of structural identifiability is typical of overly complex models, that is, those containing more parameters than can be supported by the evidence even in the utopian case of perfect measurements. As a consequence, the estimated values of structurally unidentifiable parameters are biologically meaningless. 
Structural identifiability is considered a prerequisite and necessary condition for the success of any parameter estimation procedure \cite{raue2014comparison,distefano2014dynamic,villaverde2016}.
Furthermore, the use of a structurally unidentifiable model for predicting the time course of system variables that cannot be directly measured can produce wrong results \cite{cobelli1980parameter}. Hence, the usefulness of a model for obtaining biological insight is compromised if structural identifiability is not taken into account.

%%% Paper organization:
In this paper we begin by illustrating the correspondence between the definition of dynamical compensation (DC1) and structural unidentifiability \cite{sontag2016dynamic,villaverde2017dynamical}, using the same case studies as in the original publication \cite{karin}. Then we suggest a more complete definition (DC2) drawing from ideas implicit in \cite{karin}.
After discussing the implications that lack of structural identifiability has in biological modelling, we argue that it is important to assess, and ideally enforce, the structural identifiability of a model before using it to extract insights about the corresponding biological system. 
Given that structural identifiability is a desirable model property, we enquire whether it is possible to reconcile the concept of dynamical compensation with it. We provide a positive answer by suggesting an alternative definition of dynamical compensation (DC-Id) which does not necessarily imply lack of structural identifiability, and preserves the intended meaning of the DC concept. Using for illustrative purposes one of the circuits proposed by Karin et al. \cite{karin}, we explore different modelling choices, show how they affect the identifiability of the model, and propose feasible alternatives that lead to structurally identifiable models with dynamical compensation. 
%Finally, we conclude the paper by arguing for the need of drawing a connection between newly coined concepts and the existing theoretical results available in the different communities working in biological modelling.

%Parameter estimation in biological models \cite{van2006dynamic,jaqaman2006linking,ashyraliyev2008systems,banga2008parameter}

%%%%%%%%%%%%%%%%%%%%%%%%%%%%%%%%%%%%%%%%%%%%%%%%%%%%%%%%%%%%%%%%%%%%%%%%%%%%%%%%%%%%%%%%%%%%%%%
%%%%%%%%%%%%%%%%%%%%%%%%%%%%%%%%%%%%%%%%%%%%%%%%%%%%%%%%%%%%%%%%%%%%%%%%%%%%%%%%%%%%%%%%%%%%%%%
\section*{Results}\label{sec:results}
%%%%%%%%%%%%%%%%%%%%%%%%%%%%%%%%%%%%%%%%%%%%%%%%%%%%%%%%%%%%%%%%%%%%%%%%%%%%%%%%%%%%%%%%%%%%%%%
%%%%%%%%%%%%%%%%%%%%%%%%%%%%%%%%%%%%%%%%%%%%%%%%%%%%%%%%%%%%%%%%%%%%%%%%%%%%%%%%%%%%%%%%%%%%%%% 

%%%%%%%%%%%%%%%%%%%%%%%%%%%%%%%%%%%%%%%%%%%%%%%%%%%%%%%%%%%%%%%%%%%%%%%%%%%%%%%%%%%%%%%%%%%%%%%
\subsection*{The original definition of dynamical compensation (DC1) is equivalent to structural unidentifiability}\label{sec:equiv}

It was noted in the Introduction that, according to its original definition (DC1), dynamical compensation is equivalent to structural unidentifiability.
Here we demonstrate this equivalence by interrogating the structural identifiability of the parameters of the four case studies presented in \cite{karin}. They are the circuits shown in Figure 1; they model possible regulatory mechanisms and are described by ordinary differential equations (ODEs). Since some of these models are nonlinear (C and D) and, in one case, also non-rational (circuit D), we used the STRIKE-GOLDD tool \cite{villaverde2016}, which is capable of analysing the structural identifiability of this type of systems. More details about the models and their analyses are provided in the Methods and Models section. 

\begin{figure*}[t]
	\begin{center}
		\includegraphics[width=1\linewidth]{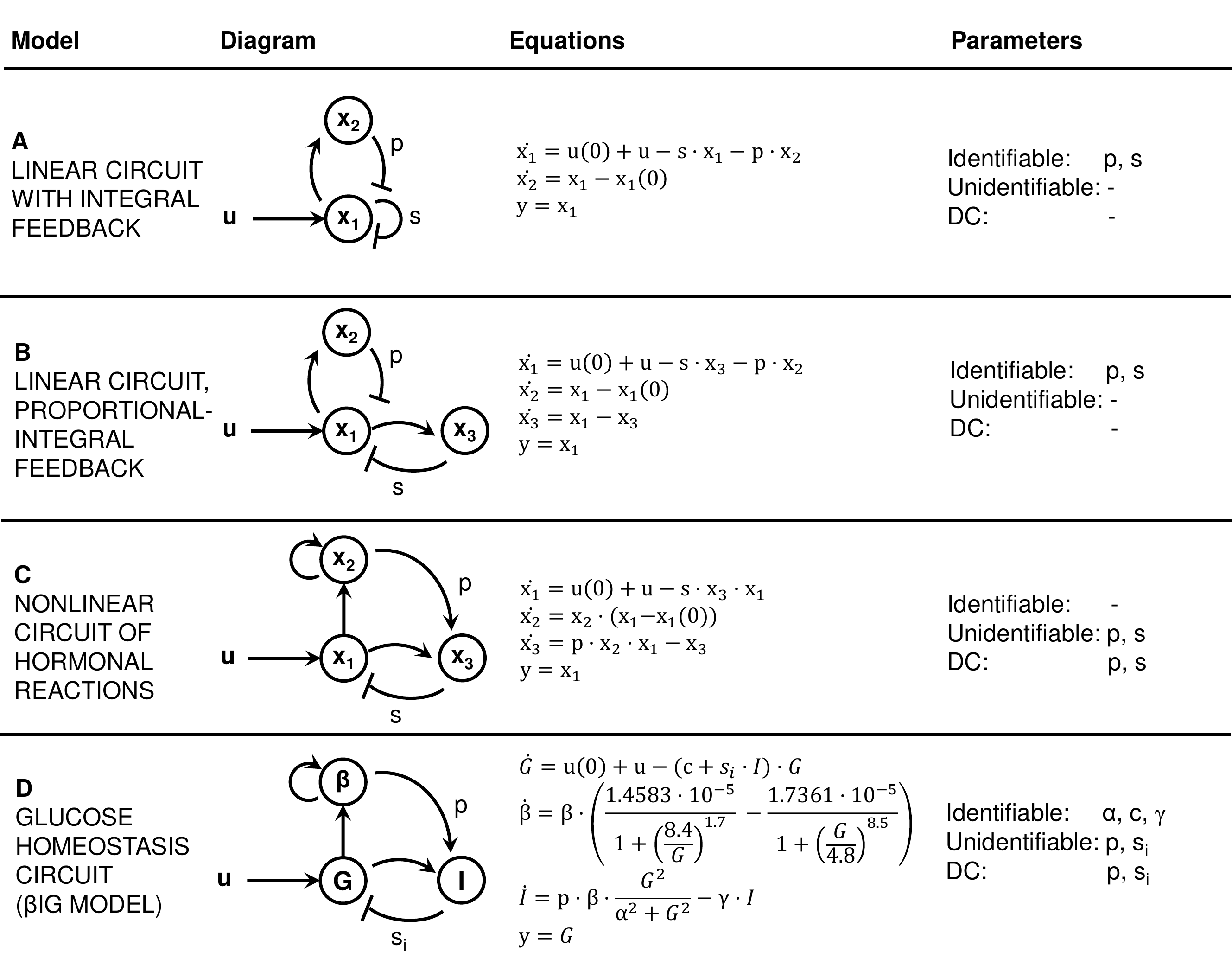}
		\caption{Physiological systems used as case studies by Karin et al. \cite{karin}. %The four circuits shown here are deterministic dynamic models described by ordinary differential equations (ODEs). 
        Parameters $p$ and $s$ (or $s_i$) represent gain constants of the feedback loops present in the circuits.}\label{fig:4circuits}  
	\end{center}	
\end{figure*} 

The parameters ($p, s$) of the two models exhibiting dynamical compensation -- i.e. the hormone circuit of Fig. 1C and the $\beta$IG model of Fig. 1D -- are structurally unidentifiable, while in the models that have exact adaptation but not dynamical compensation -- i.e. the ones shown in Fig. 1A and 1B -- those parameters are identifiable. Likewise, parameters ($\alpha, c, \gamma$) in the $\beta$IG model are structurally identifiable, as might be expected given that the model does not have dynamical compensation with respect to them.

%%%%%%%%%%%%%%%%%%%%%%%%%%%%%%%%%%%%%%%%%%%%%%%%%%%%%%%%%%%%%%%%%%%%%%%%%%%%%%%%%%%%%%%%%%%%%%%
\subsection*{The meaning of dynamical compensation and an alternative definition (DC2)}

The DC1 definition examined so far is the one explicitly provided by Karin et al. \cite{karin}. 
As explained above, DC1 does not mention explicitly certain aspects whose omission can lead to confusion, and in fact, it can be considered as a rephrasing of the structural unidentifiability property \cite{sontag2016dynamic,villaverde2017dynamical}.
However, the concept of dynamical compensation was not introduced with the aim of describing the same issue as structural unidentifiability. 
Instead, it was purported to describe a different phenomenon, specifically relevant for the regulation of physiological systems.
To clarify the intended meaning of dynamical compensation in the context it was proposed, we use the $\beta$IG model of Figure 1D as an example.
This model describes a glucose homeostasis mechanism where $\beta$ stands for the beta-cell functional mass, $I$ for insulin, and $G$ for glucose.

\begin{figure*}[t]
	\centering
	\includegraphics[width=1\linewidth]{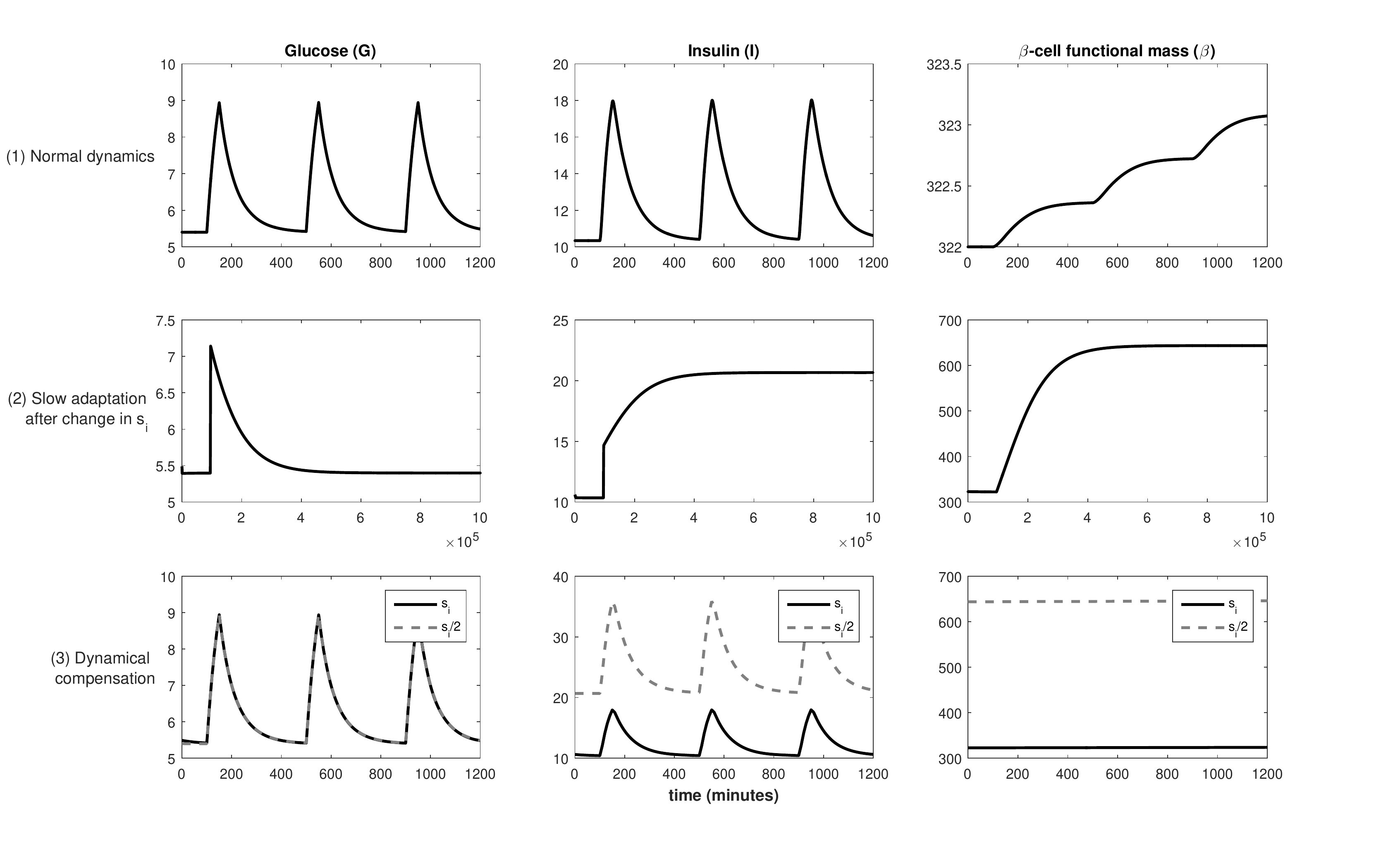}
	\caption{Illustration of the phenomenon of dynamical compensation in a physiological circuit: the $\beta$IG model. The upper row shows the normal behaviour of the system for a given value of the $s_i$ parameter. The second row shows the evolution of the steady state after a change in the value of $s_i$. After a long adaptation period, which can take months, a new steady state is reached. Then, as shown in the third row, the response of the glucose concentration for the new parameter value is the same as the initial one (this does not happen for insulin and $\beta$-cell mass).}  
	\label{fig:plots}	
\end{figure*}

The time evolution of its three states in typical scenarios is shown in Figure \ref{fig:plots}.
The first row describes the behaviour after a pulse in glucose, for example after meals.
Both glucose and insulin concentrations reach peaks shortly after the meals, and in a few hours they return to their normal levels (steady state).

The second row describes what happens if the value of a parameter, insulin sensitivity ($s_i$), is changed. Specifically, the figure represents the case in which $s_i^{new}=0.5s_i^{old}$.
There is a slow adaptation of the system's steady state, which can take months, as seen in the figure. After this period the system has adapted to a new steady state: for glucose concentration it remains the same as the initial one, while the values of insulin concentration and $\beta$-cell mass are doubled.

The third row illustrates the phenomenon of DC itself: after the adaptation to a new steady state has occurred, the output of the system (from the new steady state, and with the new value of insulin sensitivity) as a response to a pulse in glucose is \textit{the same as before the parameter change} (from the old steady state and the old parameter value). 
Note that only the glucose dynamics remains unchanged; for insulin and $\beta$-cell mass there is a scaling.  

In light of this behaviour, the following alternative definition of dynamical compensation (DC2) may be deduced from a detailed reading of the original paper by Karin et al.:

\textbf{DC2 definition of dynamical compensation:}
``Consider a model of a dynamical system with an input $u(t)$, a set of states $x(t)$, and an output $y(t,s))$, such that $s > 0$ is a \textit{known} parameter. The system is initially at a steady state $x(0)=\xi$ with $u(0) = 0$. The dependence of the output on the initial steady state is denoted by $y(t,s,\xi)$.
 Dynamical compensation (DC) with respect to $s$ is that for any parameter values $s_1$, $s_2$, for any time-dependent input $u(t)$, and for two different initial steady states, $\xi_1 \neq \xi_2$, the output of the system does not depend on $s$, that is, $y(t,s_1, \xi_1) = y(t,s_2, \xi_2)$.''

\begin{figure*}[t]
	\centering
	\includegraphics[width=1\linewidth]{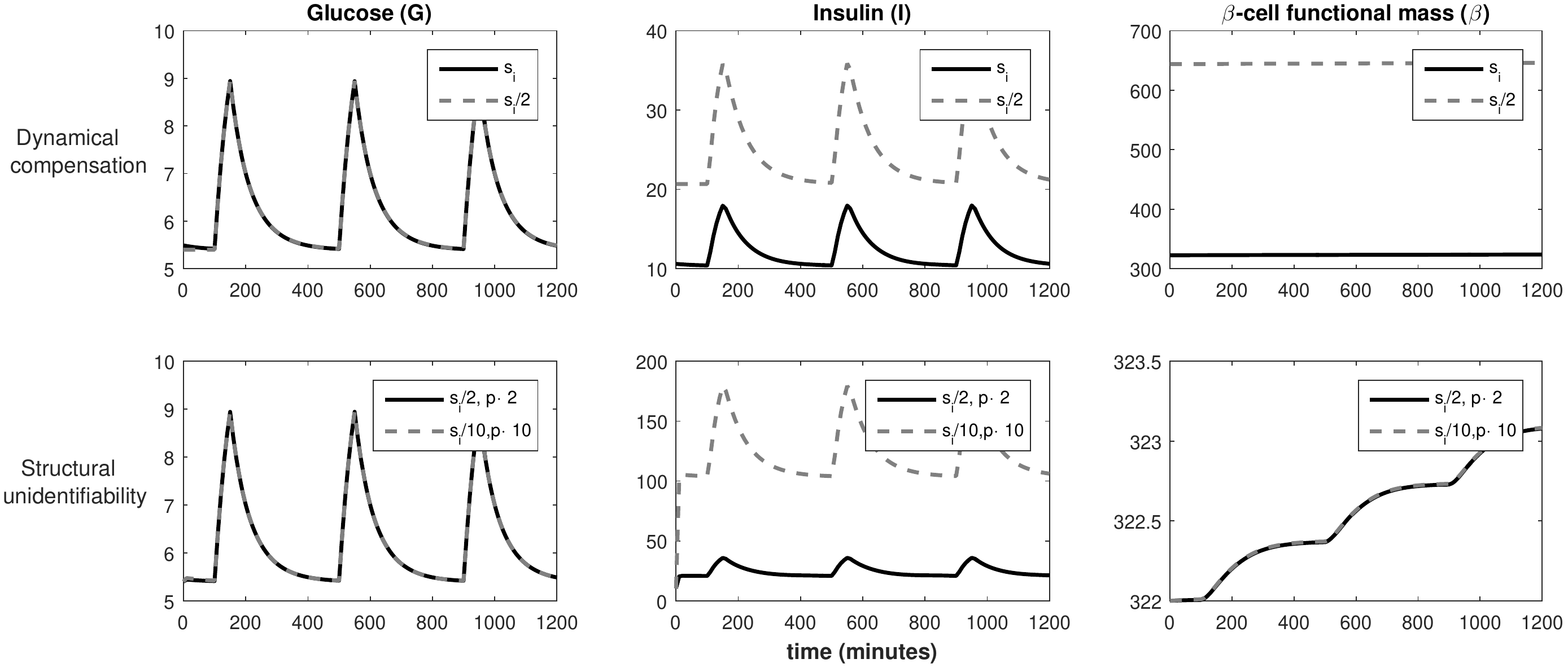}
	\caption{Dynamical compensation and structural unidentifiability in the $\beta$IG model. The first row reproduces the last row of Fig. \ref{fig:plots} and illustrates the phenomenon of dynamical compensation: after the system has adapted to the new value of $s_i$, the time-evolution of the glucose concentration (G) for the new value of ($s_i/2$) is the same as it was with the old value before adaptation ($s_i$). The second row illustrates the phenomenon of structural unidentifiability: the time-evolution of the glucose concentration (G) is the same for any value of the parameter $s_i$, as long as any deviations from the original value are compensated by changes in the parameter $p$. Note that, since the upper and lower plots of G are identical, if glucose is the only measured quantity both phenomena cannot be distinguished. However, the behaviour of the other state variables (I, $\beta$) can be very different, as can be noticed from the second and third columns.}  
	\label{fig:plots3}	
\end{figure*}

We have tried to keep this new definition, DC2, as similar as possible to DC1, and used the same notation.
The DC2 definition of dynamical compensation makes it different from structural unidentifiability. However, it assumes that all model parameters are known, which is usually not realistic. Why is this requirement of known parameters necessary? Let us illustrate this point with Fig. \ref{fig:plots3}. It shows in its first row the aforementioned example of dynamical compensation, which has already been discussed (the first row in  Fig. \ref{fig:plots3} is the same as the third one in \ref{fig:plots}).
Now, let us assume that the parameters  $p, s_i$ are unknown, to see the role played by structural unidentifiability. This is shown in the second row of Fig. \ref{fig:plots3}.
It can be seen that different values of $s_i$ result in the same dynamic behaviour of glucose concentration, as long as the change in $s_i$ is compensated by a coordinated change in $p$. 
If, as suggested in \cite{karin}, glucose is the only measured variable, $p, s_i$ are structurally unidentifiable: their values cannot be determined, because there is an infinite number of possible combinations of values that yield the same output. 
This can be problematic, because choosing wrong values for $p, s_i$ results in incorrect predictions of the concentration of insulin, as can be observed in the second column.
In fact, values of $p, s_i$ that yield the same curve of glucose can correspond to totally unrealistic curves of insulin. 
Importantly, since there is only one measured output (glucose), dynamical compensation cannot be distinguished from structural unidentifiability. For this reason, dynamical compensation can only be claimed if structural unidentifiability can be ruled out. This is the reason for enforcing known parameters in the DC2 definition.

%%%%%%%%%%%%%%%%%%%%%%%%%%%%%%%%%%%%%%%%%%%%%%%%%%%%%%%%%%%%%%%%%%%%%%%%%%%%%%%%%%%%%%%%%%%%%%%
\subsection*{Implications of structural unidentifiability}\label{implications}

The fact that a model is unidentifiable is important because, after five decades of research, it is now well understood that lack of structural identifiability is the result of choosing an inappropriate model structure for the available measurable variables (or variables that can be directly observed) \cite{walter1997identification,distefano2014dynamic}. When understood in this way, structural unidentifiability can be avoided or surmounted in at least three ways: (i) by reducing the number of parameters or changing their definition, (ii) by increasing the number of measured variables, if possible, or (iii) by determining the unidentifiable parameters in some alternative way, e.g. by direct measurements. 

Strategy (i) entails reformulating the model to remove redundant parameters, for example, by grouping several non-identifiable parameters into a single identifiable one. Perhaps the simplest example would be the merging of two parameters that multiply each other into a single one, i.e. $p_{\text{new}}= p_1\times p_2$. 
Such relationships can be revealed systematically by performing a structural identifiability analysis.

Strategy (ii) can be illustrated with the ``$\beta$IG'' model: if it were possible to measure all its three states instead of only glucose, all the parameters in the $\beta$IG model would become structurally identifiable. In other words, while the effect on the glucose concentration (G) of a change in $p$ can be compensated by changing $s_i$, this does not happen for the insulin concentration (I). This point will be discussed in more detail in the following section.

Strategy (iii) was applied for example by Watson et al. \cite{watson2011new}. After determining that two parameters in a homeostatic model were structurally unidentifiable, they decided to measure one of them by means of a tracer experiment and to calculate an estimate of the other using a steady state assumption. We will discuss a similar assumption in the context of the $\beta$IG model in the next section.

Strategies (ii) and (iii) demonstrate how measurements and data can directly inform modelling decisions.  More generally, structural identifiability analysis can inform expectations about how precisely a model can be defined, given measurements and data.  For example, if the state of a system changes very little when a parameter varies, the system is said to be robust or insensitive to variations in that parameter \cite{distefano2014dynamic}. Speaking in terms of identifiability, this situation corresponds to poor \textit{practical} (or \textit{numerical}) \textit{identifiability}. Although the value of the parameter has some influence on the model output, its effect is too small to allow for its precise determination due to limitations in data quantity and/or quality\cite{jacquez1985numerical,walter1997identification}. In contrast, when the sensitivity of the model output to a parameter is exactly zero -- as implied by DC1 -- it corresponds to lack of \textit{structural identifiability}. In this case, the value of the parameter has no influence at all on the model output. %This ``unreasonable elasticity'' indicates that the parameter is not meaningful; ideally, it should be removed and the model should be modified, as explained above.

As recently stressed by Janz\'en et al. \cite{janzen2016}, the danger of inadvertently using a structurally unidentifiable model is that the biological interpretations of its parameters are not valid, which may lead to wrong conclusions; furthermore, any predictions involving unmeasured states ``may be meaningless if the parameters directly or indirectly related to those states are unidentifiable''  \cite{janzen2016}. 
This can be illustrated with the $\beta$IG model, as seen in the second row of Fig. \ref{fig:plots3}: if we try to estimate the $p,s_i$ parameters from glucose (G) measurements, we will not be able to recover their true values, because they are structurally unidentifiable: there is an infinite number of combinations of their values that yield the same glucose profile. This, in turn, means that we cannot use the model to predict the time-course of insulin concentration (I), which is an unmeasured state. As seen in the lower plot of the second column, the predictions of insulin can be very different depending on the pair of $p,s_i$ values used. 

While structural identifiability analysis does not inform us of the plausibility of a biological mechanism, it can warn us that a particular model formulation is not adequate for describing the mechanism. In the words of Bellman and {\AA}str{\"o}m \cite{bellman1970structural}, the concept of structural identifiability ``is useful when answering questions such as: To what extent is it possible to get insight into the internal structure of a system from input-output measurements? What experiments are necessary in order to determine the internal couplings uniquely?'' 

Physiological models have often been the subject of structural identifiability analyses in the past; these studies may serve to illustrate why asking these questions make one’s interpretation of the model more rigorous. For example, DiStefano III \cite{distefano2014dynamic} reported identifiability issues in models of blood glucose control such as the classic one by Bolie \cite{Bolie}, among others. The Bolie model (see the Methods section for details) has two state variables, glucose and insulin, and five parameters. None of its parameters are identifiable if plasma glucose concentration is the only measured variable (let us refer to this model, which includes one observable, as ``Bolie A''). However, if insulin concentration is also measured (``Bolie B''), all of them become identifiable. If it is not possible to measure insulin, a structurally identifiable model can be obtained by reducing the number of unknown parameters to three, e.g. by fixing the values of two of them (``Bolie C''). The search for DC1 in model ``Bolie A'' could lead to the conclusion that the proposed mechanism for glucose concentration is infinitely robust to changes in the values of its parameters, and that this represents a biologically relevant feature. However, the same mechanism becomes structurally identifiable when modelled as ``Bolie B'' or ``Bolie C''. This example remarks that structural unidentifiability is not a property of a given mechanism, but a consequence of modelling choices. 

It should be noted that structural identifiability has close links with other properties, namely observability and controllability \cite{distefano1977relationships}, which provide important information about a dynamic model. Observability determines whether it is possible to reconstruct the internal state of a model by observing its output, while controllability determines the states that the system can reach by manipulating its input. Identifiability and observability are tightly related; in fact, structural identifiability analysis can be recast as observability analysis by considering the model parameters as constant state variables. This is the approach adopted by the STRIKE-GOLDD toolbox \cite{villaverde2016} used in our analyses. In turn, observability and controllability are usually considered as dual concepts. These three distinct but not unrelated properties -- identifiability, observability, and controllability -- can be analysed for general nonlinear ODE models using differential geometric approaches, which are described e.g. in \cite{Vidyasagar1993,sontag1982mathematical}.

Finally, it should be mentioned that in a realistic parameter estimation  scenario it is also necessary to take into account limitations introduced by the quantity and quality of the available data. This is the related topic of \textit{practical} or \textit{numerical} identifiability, which aims at quantifying the uncertainty in the estimated parameter values that results not only from the model structure but also from data limitations \cite{jacquez1985numerical,walter1997identification,distefano2014dynamic,villaverde2016identifiability}. 

%%%%%%%%%%%%%%%%%%%%%%%%%%%%%%%%%%%%%%%%%%%%%%%%%%%%%%%%%%%%%%%%%%%%%%%%%%%%%%%%%%%%%%%%%%%%%%%
\subsection*{Dynamical compensation in realistic scenarios: DC-Id}\label{sec:dc2}

In reality, biological models almost always have a number of unknown parameters, whose values must be determined before the model can be used in practical applications. 
In this context the following question naturally arises: 
how does the behaviour described by the concept of dynamical compensation relate with structural identifiability of the parameters in the model? 
As we have already mentioned structural unidentifiability was shown to be equivalent to the (original explicit definition of) dynamical compensation \cite{sontag2016dynamic,villaverde2017dynamical}. Is this, then, the end of the question? Are systems with dynamical compensation ``doomed'' to be represented by structurally unidentifiable models, thus potentially limiting the biological insight that can be extracted from them?

We claim here that this is not necessarily the case, provided that we reformulate the definition of dynamical compensation. To show this, let us examine in more detail the structural identifiability of a system with dynamical compensation, the $\beta$IG model of Fig. 1D.
We analysed the structural identifiability of this model in its original formulation earlier in this paper, when we showed that the original definition of dynamical compensation (DC1) is equivalent to structural unidentifiability. In that section we showed that, when its five parameters $(p,s_i,\gamma,c,\alpha)$ are considered unknown and plasma glucose concentration (G) is the only available measurement, the $\beta$IG model is structurally unidentifiable. More specifically, the two parameters that exhibit dynamical compensation ($p,s_i$) are unidentifiable, while the remaining three are identifiable.
 
Let us now see the results of such analysis when we change key aspects of the model, while preserving its dynamics.
The two main choices we can play with are: (i) which parameters of the model are considered unknown, and therefore need to be estimated; and (ii) which measurements are possible. 
Regarding the first choice (i), we analyse not only the five-parameter case considered by Karin et al., but also other representative scenarios: when the unknown parameters are $\{s_i,\gamma,c,\alpha\}$ (i.e., all but $p$), when they are $\{p,s_i\}$, and when there is only one unknown, $s_i$. 
The second choice (ii) defines the output function of the model. While in general the output can be any function of the states, typically it consists of a subset of the states. In the version of the $\beta$IG model used by Karin et al. \cite{karin} the only measured variable was glucose concentration (G). Here we consider all the possibilities, to assess the consequences of measuring every possible combination of the three state variables of the model: glucose (G) and insulin (I) concentrations, and beta-cell mass ($\beta$). 
Let us consider that all three measurements could be in principle feasible (although in principle it seems much easier to measure G and I than $\beta$). 
The set of 28 alternative model configurations and the corresponding results of the structural identifiability analysis are summarized in Table \ref{tab:si_results}.

\begin{table}[ht]
	\centering
	\begin{tabular}{c|c|c|c|c|}
		\cline{2-5}		
		& \multicolumn{4}{c|}{  \cellcolor{gray}\textbf{Unknown parameters}}\\
		\hline
		\rowcolor{gray}
		\multicolumn{1}{|c|}{\textbf{Outputs}} & $\{\alpha,\gamma,c,p,s_i\}$  & $\{p,s_i\}$  & $\{\alpha,\gamma,c,s_i\}$& $s_i$ \\
		\hline
		\multicolumn{1}{|c|}{G}                & $\{p,s_i\}$                  &  $\{p,s_i\}$ & $s_i$                    & $s_i$  \\
		\multicolumn{1}{|c|}{$\beta$}          & $\{p,s_i\}$                  &  $\{p,s_i\}$ & -  &  -      \\
		\multicolumn{1}{|c|}{I}                & $p$                          &  $p$         & -  &  -      \\
		\hline
		\multicolumn{1}{|c|}{G,I}              & $p$                          &  $p$         & -  &   -     \\
		\multicolumn{1}{|c|}{G,$\beta$}        & $\{p,s_i\}$                  &  $\{p,s_i\}$ & -  &   -     \\
		\multicolumn{1}{|c|}{I,$\beta$}        &   -                          &     -        & -  &    -    \\
		\hline       
		\multicolumn{1}{|c|}{$\beta$,I,G}      &   -                          &     -        & -  &   -      \\
		\hline
	\end{tabular}
	\caption{\label{tab:si_results}\textbf{Structurally unidentifiable parameters for different configurations of the $\beta$IG model.} The choice of measured outputs and parameters considered unknown determines the identifiability of the remaining parameters (those considered to be unknown). Four representative choices of parameters are studied: (i) with all the model parameters $\{\alpha,\gamma,c,p,s_i\}$ considered unknown, (ii) with the two parameters $\{p,s_i\}$ that may exhibit dynamical compensation considered unknown, (iii) with all but $p$ unknown, and (iv) with only one parameter, $s_i$, considered unknown.}
\end{table}

From these results we highlight the following observations:

\begin{itemize}
	\item The three parameters that do not have dynamical compensation, $\{\alpha,\gamma,c\}$, are always identifiable in this model, and knowing them does not change the identifiability of those that have dynamical compensation, $\{p,s_i\}$ (note that the second and third columns in Table \ref{tab:si_results} are identical, and so are the fourth and fifth).
	\item If glucose (G) is the only output, it is always impossible to identify $s_i$, even if all the remaining parameters are known.
	\item If other state variables can be measured, measuring also glucose does not improve the structural identifiability. %(although it will probably improve \textit{practical} identifiability). \textcolor{red}{mention pract identif here?}
	\item The ``cheapest'' ways of obtaining an identifiable model are:
	\begin{enumerate}[label=(\Alph*)]
		\item If $p$ is assumed to be known, or can be estimated in some way other than from input-output data, we obtain a structurally identifiable model by measuring only insulin concentration (we could measure $\beta$-cell mass instead of insulin, but this seems to be technically more difficult, or maybe unfeasible).
		\item If all parameters including $p$ are unknown: since the $\beta$-cell mass changes very little (as can be seen in Fig. \ref{fig:plots3}), we can assume it constant and use as an estimate of it a single measurement obtained in the past. With this assumption it suffices to monitor the insulin concentration to obtain an identifiable model.
	\end{enumerate}
\end{itemize}

It can be noticed that there is substantial variability in the identifiability results depending on the modelling choices.
We remark however that these choices do not alter the dynamics: the dynamic behaviour of the system is the same for all the different configurations above.
How do these different configurations affect dynamical compensation?
Both the original definition of dynamical compensation (DC1) and the second one (DC2) consider single-output models.  
Specifically, Karin et al. demonstrated that the $\beta$IG model has dynamical compensation in glucose concentration (G) with respect to the $\{p,s_i\}$ parameters. As can be seen in the first row of Table \ref{tab:si_results}, both parameters are structurally unidentifiable when G is the only output.
To break this correspondence between dynamical compensation and structural unidentifiability we might interpret the ``output'' in the DC definition to be multi-dimensional. If we measure not only glucose but also other state variable(s) we can make $\{p,s_i\}$ identifiable, at least in some cases. However, this would entail losing the dynamical compensation property, because it holds for glucose but not for the other state variables that may form the output of the system.
Thus, additional precisions should be incorporated into our working definition of dynamical compensation in order to make it describe a meaningful systemic property without being equivalent to structural unidentifiability.
In light of this, we propose the following definition of dynamical compensation, which we call DC-Id:

\textbf{DC-Id definition of dynamical compensation:}
``Consider a nonlinear time-invariant dynamic system modelled as a structure $M$ with the following equations: 

\begin{equation}\label{eq:M}
M:\left\{%
\begin{array}{lll}
\dot{x}(t) & = & f(x(t),p,u(t))\\
y(t) & = & h(x(t),p)\\
x_0 & = & \xi(p)\\
\end{array}%
\right.
\end{equation}

\noindent where $f$ and $h$ are vector functions, $p\in\mathbb{R}^q$ is a real-valued vector of parameters, $u\in\mathbb{R}^r$ is the input vector, $x\in\mathbb{R}^n$ the state variable vector, and $y\in\mathbb{R}^m$ the output or observables vector. 
The parameters $p$ can be known or unknown constants. The system is initially at a steady state $\xi$, with $u(0) = 0$. 
The initial state is denoted as $x_0$.
The dependence of the $i^{th}$ output $y_i(t)$ on the initial steady state and on a particular value of a parameter $p_i \subset p$ can be made explicit by writing it as $y_i(t|p_i=k,x_0=\xi)$.
Then, dynamical compensation (DC) of a particular model output $y_i \subset y$ with respect to a parameter $p_i \subset p$ is that, for any two values of $p_i$ ($k_1$ and $k_2$) and for two different initial steady states ($\xi_1 \neq \xi_2$), the output $y_i$ does not depend on $p_i$, that is, $y_i(t|p_i=k_1, x_0=\xi_1) = y_i(t|p_i=k_2, x_0=\xi_2)$, for any time-dependent input $u(t)$.''

This new definition effectively distinguishes the phenomenon of dynamical compensation from the structural unidentifiability property, by explicitly acknowledging that it applies to a subset (possibly only one) of the model outputs and to a subset of the parameters. 
When applied to the different model configurations of Table \ref{tab:si_results}, the DC-Id definition yields that there is indeed dynamical compensation for the G output (glucose concentration) with respect to the $\{p,s_i\}$ parameters in all cases, and not only for the unidentifiable ones.

%%%%%%%%%%%%%%%%%%%%%%%%%%%%%%%%%%%%%%%%%%%%%%%%%%%%%%%%%%%%%%%%%%%%%%%%%%%%%%%%%%%%%%%%%%%%%%%
\section*{Discussion}\label{sec:conclusions}

The ability to exhibit robust behaviour in the face of changing external conditions is a remarkable feature of many biological processes.  
It is now widely accepted that negative feedback plays a central role in biological phenomena such as homeostasis. Feedback mechanisms are capable of rendering a system robust to a wide range of external disturbances (or, in other words, they can compensate for changes in environmental conditions). %without the need for changing the model parameters. 
Dynamic modelling is a powerful tool for analysing possible regulatory mechanisms, such as feedback circuits, and gaining insight about the corresponding biological systems. However, modelling artefacts such as those arising from structural unidentifiability may lead to wrong conclusions if not properly accounted for. 
The reason is that a system's dynamics determines what it can \textit{do}, while its identifiability (and observability) determines what we can \textit{know} about it. Hence deficiencies in identifiability may lead to wrong reconstructions of a system's behaviour. 
Therefore, the structural identifiability of a model should be assessed before the model is used to extract insights about (and ultimately understand) the corresponding biological system.

For this reason we believe that it is necessary and useful to draw an explicit connection between the newly coined concept of dynamical compensation and the existing literature and theory on structural identifiability, especially taking into account that parameter identification is an ubiquitous need in biological modelling. 
With this aim, in the present work we have shown that the absence of such connection in the paper that introduced dynamical compensation \cite{karin} led to an ambiguous definition of the concept, which we have termed DC1. The fact that DC1 is essentially equivalent to structural unidentifiability when examined from the viewpoint of model identification \cite{sontag2016dynamic,villaverde2017dynamical} is a source of potential confusion: it opens the door to (i) interpreting as dynamical compensation what might be a case of structural unidentifiability, and to (ii) inadvertently using structurally unidentifiable models. Such use has associated risks, which we have discussed in this work. 
It is possible to deduce from a detailed reading of the original paper \cite{karin} an alternative definition of dynamical compensation, which we have called DC2. While DC2 can be considered implicit in \cite{karin}, we have stated it explicitly in the present paper and have shown that it removes some ambiguities of DC1. However, neither DC1 nor DC2 are appropriate for realistic modelling scenarios, in which it is necessary to estimate the values of parameters from input-output data. 
To address this issue we have proposed a modification of the definition of dynamical compensation which can be used in such cases.  
Our new definition, termed DC-Id, captures the biological meaning of the dynamical compensation phenomenon, which is the invariance of the dynamics of certain state variables of interest with respect to changes in the values of certain parameters. 
But, additionally, it includes precisions that make it distinct from structural unidentifiability, even in the context of parameter identification -- that is, when it is necessary to determine the values of the model parameters.

In summary, we have discussed three alternative definitions of dynamical compensation, which we have termed DC1, DC2, and DC-Id. We have seen that DC1 is equivalent to structural unidentifiability, DC2 is different but of limited utility, and the proposed DC-Id is unambiguous and generally applicable. 
We see the discussion held in the present paper and the resulting clarification as an example of the gains that can be obtained by exchanging more notes among the different communities working in systems biology, which we have advocated elsewhere \cite{villaverde2014reverse}. Such an exchange of notes increases researchers' awareness of community-specific knowledge and is useful for avoiding potential misconceptions.

%%%%%%%%%%%%%%%%%%%%%%%%%%%%%%%%%%%%%%%%%%%%%%%%%%%%%%%%%%%%%%%%%%%%%%%%%%%%%%%%%%%%%%%%%%%%%%%
%%%%%%%%%%%%%%%%%%%%%%%%%%%%%%%%%%%%%%%%%%%%%%%%%%%%%%%%%%%%%%%%%%%%%%%%%%%%%%%%%%%%%%%%%%%%%%%
\section*{Methods and Models}\label{sec:methods}
%%%%%%%%%%%%%%%%%%%%%%%%%%%%%%%%%%%%%%%%%%%%%%%%%%%%%%%%%%%%%%%%%%%%%%%%%%%%%%%%%%%%%%%%%%%%%%%
%%%%%%%%%%%%%%%%%%%%%%%%%%%%%%%%%%%%%%%%%%%%%%%%%%%%%%%%%%%%%%%%%%%%%%%%%%%%%%%%%%%%%%%%%%%%%%%

%%%%%%%%%%%%%%%%%%%%%%%%%%%%%%%%%%%%%%%%%%%%%%%%%%%%%%%%%%%%%%%%%%%%%%%%%%%%%%%%%%%%%%%%%%%%%%%
\subsection*{Mathematical notation}
Following the convention usually adopted when modelling dynamical systems, we use: $x$ to refer to state variables, $u$ for inputs, $y$ for outputs, and $p$ for parameters. States, inputs, and outputs are in general time-varying, while parameters are constants. We consider models described by ordinary differential equations of the following general form:

\begin{equation}\label{eq:model_x}
M:\left\{%
\begin{array}{lll}
\dot{x}(t) & = & f(x(t),p,u(t))\\
y(t) & = & h(x(t),p)\\
x_0 & = & x(p)\\
\end{array}%
\right.
\end{equation}

where $f$ and $h$ are analytic vector functions of the states and parameters, which are in general nonlinear (linear models are a particular case). For ease of notation we will omit the dependence of $f$ and $h$ on $p$, and denote initial values of state variables or inputs as $x_0=x(0)$ and $u_0=u(0)$, respectively. 
We will also generally drop the time dependence, i.e. we will write $x$ instead of $x(t)$, and so on.
Note that by ``model structure'' we refer not only to the dynamic equations ($\dot{x}$) but also to the definition of the observation function, or set of measured model outputs ($y$), and the known input variables ($u$).

%%%%%%%%%%%%%%%%%%%%%%%%%%%%%%%%%%%%%%%%%%%%%%%%%%%%%%%%%%%%%%%%%%%%%%%%%%%%%%%%%%%%%%%%%%%%%%%
\subsection*{Nonlinear observability}
Among the existing approaches for structural identifiability (SI) analysis, we adopt one that considers SI as a generalization of observability -- the property that allows reconstructing the internal state ($x$) of a model from observations of its outputs ($y$). If a model is observable there is a unique mapping from $y$ to $x$, and two different states will lead to two different outputs. Observability is a classic system-theoretic property introduced by Kalman for linear systems, and extended to the nonlinear case by Hermann and Krener \cite{hermann1977nonlinear}, among others. It can be studied with a differential geometry approach, as described in the remainder of this subsection. A thorough treatment of this matter can be found in \cite{Vidyasagar1993,sontag1982mathematical}.

Observability analysis determines if the mapping from $y$ to $x$ is unique by analysing the expression of $y=h(x)$ and its derivatives. This is done by constructing an observability matrix that defines this mapping, and then calculating its rank. If the matrix is full rank then there is a one-to-one correspondence between outputs and states, and the system is observable. If not, the same output can be produced by several state vectors, and the system is unobservable. In the nonlinear case, the observability matrix is built using Lie derivatives. The Lie derivative of $h$ with respect to $f$ is:

\begin{equation}
L_f h(x) = \frac{\partial h(x)}{\partial x}f(x,u)
\end{equation}

Higher order Lie derivatives can be calculated from the lower order ones as:

\begin{align}
\begin{array}{rcl}
L_f^2 h(x) & = & \frac{\partial L_f h(x)}{\partial x}f(x,u) \\
& \cdots & \\
L_f^i h(x) & = & \frac{\partial L_f^{i-1} h(x)}{\partial x}f(x,u)
\end{array}
\end{align}

The nonlinear observability matrix can be written as:

\begin{align}\label{nonlinobs}
{\mathcal O}(x) = \left(  
\begin{array}{c}
\frac{\partial }{\partial x}h(x)           \\
\frac{\partial }{\partial x}(L_f h(x))     \\
\frac{\partial }{\partial x}(L_f^2 h(x))   \\
\vdots       \\
\frac{\partial }{\partial x}(L_f^{n-1}h(x))\\      
\end{array}
\right)
\end{align}

\noindent where $n$ is the dimension of the state vector $x$. We can now formulate the \textit{Observability Rank Condition (ORC)} as follows:
if the system given by (\ref{eq:model_x}) satisfies $\text{rank}({\mathcal O}(x_0)) = n$, where ${\mathcal O}$ is defined by (\ref{nonlinobs}), then it is (locally) observable around $x_0$ \cite{hermann1977nonlinear}.
This condition guarantees \textit{local} observability, which means that the state $x_0$ can be distinguished from any other state in a neighbourhood, but not necessarily from distant states. The distinction between local and global identifiability is usually not relevant in biological applications.

%%%%%%%%%%%%%%%%%%%%%%%%%%%%%%%%%%%%%%%%%%%%%%%%%%%%%%%%%%%%%%%%%%%%%%%%%%%%%%%%%%%%%%%%%%%%%%%
\subsection*{Structural identifiability as generalized observability}

By considering the parameters as state variables with zero dynamics ($\dot p=0$), SI analysis can be recast as observability analysis. To this end, we augment the state vector as $\tilde{x} = \left[x,p\right]$ and write the generalized observability-identifiability matrix as:

\begin{align}\label{obsident}
{\mathcal O}_I(\tilde{x} ) = \left(
\begin{array}{c}
\frac{\partial }{\partial \tilde{x}}h(\tilde{x})             \\
\frac{\partial }{\partial \tilde{x}}(L_f h(\tilde{x}))       \\
\frac{\partial }{\partial \tilde{x}}(L_f^2 h(\tilde{x}))     \\
\vdots       \\
\frac{\partial }{\partial \tilde{x}}(L_f^{n+q-1}h(\tilde{x}))\\  
\end{array}
\right)
\end{align}

\noindent where $n$ is the dimension of the state vector $x$ and $q$ is the dimension of the parameter vector $p$. We can now state a generalized Observability-Identifiability Condition (OIC): if a system satisfies $\text{rank}({\mathcal O}_I(\tilde x_0)) = n+q$, it is (locally) observable and identifiable around the state $\tilde x_0$.

If $\text{rank}({\mathcal O}_I(\tilde x_0)) < n+q$, the model contains unidentifiable parameters (and/or unobservable states). It is possible to determine the identifiability of individual parameters because each column in $O_I$ contains the partial derivatives with respect to one parameter (or state). Thus if the matrix rank does not change after removing the $i^{th}$ column the $i^{th}$ parameter is not identifiable (if the column corresponds to a state, it is not observable).

\subsection*{Software Availability}

The software used in this paper for analysing structural identifiability is STRIKE-GOLDD (STRuctural Identifiability taKen as Extended-Generalized Observability with Lie Derivatives and Decomposition). It is a methodology and a tool for structural identifiability analysis \cite{villaverde2016} which can handle nonlinear systems of a very general class, including non-rational ones. At its core is the conception of structural identifiability as a generalization of observability. 
Since the calculation of $\text{rank}({\mathcal O}_I(\tilde x_0))$ can be computationally very demanding, even for models of moderate size, STRIKE-GOLDD includes a number of algorithmic modifications to alleviate its cost. One of them is the construction of the observability-identifiability matrix ${\mathcal O}_I$ with less than $n+q-1$ derivatives. In certain cases, this reduced matrix can suffice to establish the identifiability of the whole model; in other cases, it can at least report identifiability of a subset of parameters, even if it cannot decide on the rest. Another possibility is to decompose the model in a number of submodels, which have smaller matrices whose rank is easier to compute. More details about these and other procedures included in the methodology can be found in \cite{villaverde2016}. 

STRIKE-GOLDD is an open source MATLAB toolbox that can be downloaded from \url{https://sites.google.com/site/strikegolddtoolbox/}. 
A more complete description of the tool can be found in its user manual, which is available in the website. The code and instructions for reproducing the results reported in this paper can be found in \url{https://sites.google.com/site/strikegolddtoolbox/dc}.

\subsection*{Models}

\subsubsection*{Feedback circuits from Karin et al.}

Fig. \ref{fig:4circuits} depicts the four circuits presented by Karin et al. \cite{karin} and includes their equations.
The model depicted in Figure 1A is a linear system with integral feedback on the output variable, $y$. It corresponds to the circuit of Figure 1B from \cite{karin}.
The system depicted in Figure 1B is also linear, but has proportional-integral feedback and includes an additional state variable. It corresponds to the circuit of Figure 1C from \cite{karin}.
The nonlinear model of hormonal reactions shown in Figure 1C corresponds to the one in Figure 1D from \cite{karin}. 
Finally, the fourth case study (Figure 1D) has the same high-level diagram as the previous one (Figure 1C); however, the detailed dynamics are different.
This circuit is known as the ``$\beta$IG model'' due to its three states ($\beta$, I, and G). It describes a glucose homeostasis mechanism where $\beta$ stands for the beta-cell functional mass, $I$ for insulin, and $G$ for glucose. It corresponds to the model in Figure 2 from \cite{karin}. The presence of terms such as $(8.4/G )^1.7$ makes this system non-rational, which complicates its analysis, since many structural identifiability methods can only deal with rational models. However, it is possible to analyse non-rational systems such as this one with the STRIKE-GOLDD tool described above. 

\subsubsection*{Bolie models of blood glucose regulation}
Bolie \cite{Bolie} proposed the following model, which we have taken from \cite{distefano2014dynamic}:

\begin{equation}\label{eq:Bolie_A}
\begin{array}{lll}
\dot{q_1} & = & p_1\cdot q_1-p_2\cdot q_2 + \delta\\
\dot{q_2} & = & p_3\cdot q_2+p_4\cdot q_1\\
y & = & \frac{q_1}{V_p}\\
\end{array}%
\end{equation}

Here $x_1$ and $x_2$ are the deviations of blood plasma glucose and insulin masses from their fasting levels, $\delta$ is an impulsive injection of glucose, $V_p$ is the plasma volume, and $p_1$, $p_2$, $p_3$, and $p_4$ are parameters describing distribution and interactions of glucose and insulin. We refer to the formulation above as ``Bolie A''. 

A different version, which we call ``Bolie B'', consists of the same equations plus an additional measured variable, that is, not only plasma glucose concentration ($y_1 = \frac{q_1}{V_p}$) but also insulin ($y_2 = \frac{q_2}{V_p}$) are measured.

%\begin{equation}\label{eq:Bolie_B}
%\begin{array}{lll}
%\dot{q_1} & = & p_1\cdot q_1-p_2\cdot q_2 + \delta\\
%\dot{q_2} & = & p_3\cdot q_2+p_4\cdot q_1\\
%y_1 & = & \frac{q_1}{V_p}\\
%y_2 & = & \frac{q_2}{V_p}\\
%\end{array}%
%\end{equation}

Finally, a third variant of this model, referred to as ``Bolie C'' in the main text, is simply the same as the ``Bolie A'' described above but with only three unknown parameters ($p_3$, $p_4$, and $V_p$); that is, if $p_1$ and $p_2$ are considered as known constants.

All parameters in ``Bolie A'' are structurally unidentifiable, while in ``Bolie B'' and ``Bolie C'' they are identifiable.

%%%%%%%%%%%%%%%%%%%%%%%%%%%%%%%%%%%%%%%%%%%%%%%%%%%%%%%%%%%%%%%%%%%%%%%%%%%%%%%%%%%%%%%%%%%%%%%
%%%%%%%%%%%%%%%%%%%%%%%%%%%%%%%%%%%%%%%%%%%%%%%%%%%%%%%%%%%%%%%%%%%%%%%%%%%%%%%%%%%%%%%%%%%%%%%

%%%%%%%%%%%%%%%%%%%%%%%%%%%%%%%%%%%%%%%%%%%%%%%%%%%%%%%%%%%%%%%%%%%%%%%%%%%%%%%%%%%%%%%%%%%%%%%
%%%%%%%%%%%%%%%%%%%%%%%%%%%%%%%%%%%%%%%%%%%%%%%%%%%%%%%%%%%%%%%%%%%%%%%%%%%%%%%%%%%%%%%%%%%%%%%
\section*{Acknowledgements}
This work was supported by the Spanish Ministerio de Econom\'ia y Competitividad (and the FEDER) project "SYNBIOFACTORY" (DPI2014-55276-C5-2-R).	

%%%%%%%%%%%%%%%%%%%%%%%%%%%%%%%%%%%%%%%%%%%%%%%%%%%%%%%%%%%%%%%%%%%%%%%%%%%%%%%%%%%%%%%%%%%%%%%
%%%%%%%%%%%%%%%%%%%%%%%%%%%%%%%%%%%%%%%%%%%%%%%%%%%%%%%%%%%%%%%%%%%%%%%%%%%%%%%%%%%%%%%%%%%%%%%
\section*{Author Contributions}
Conceptualization, A.F.V. and J.R.B.; Methodology, A.F.V.; Software, A.F.V.; Writing -- Original Draft, A.F.V. and J.R.B.

\end{document}